\documentclass[review]{elsarticle}

\usepackage{lineno,hyperref}
\usepackage{latexsym}
\usepackage{graphicx}
\usepackage{amsmath,amsfonts}
\usepackage{setspace}
\usepackage[]{graphicx}
\usepackage{tocloft}
\usepackage{braket}
\usepackage{epstopdf}

\journal{Journal of \LaTeX\ Templates}

\begin{document}

\begin{frontmatter}

\title{Deterministic construction of arbitrary $W$ states with quadratically increasing number of two-qubit gates}

\author[mymainaddress]{Firat Diker\corref{mycorrespondingauthor}}

\cortext[mycorrespondingauthor]{Corresponding author}
\ead{firatdiker@sabanciuniv.edu}

\address[mymainaddress]{Faculty of Engineering and Natural Sciences, Sabanci University, Tuzla, 34956 Istanbul, Turkey}

\begin{abstract}
We propose a quantum  circuit composed of $cNOT$ gates and four single-qubit gates to generate a $W$ state of three qubits. This circuit was then enhanced by integrating two-qubit gates to create a $W$ state of four and five qubits. After a couple of enhancements, we show that an arbitrary $W$ state can be generated depending only on the degree of enhancement. The generalized formula for the number of two-qubit gates required is given, showing that an $n$-qubit $W$-state generation can be achieved with quadratically increasing number of two-qubit gates. Also, the practical feasibility is discussed regarding photon sources and various applications of $cNOT$ gates.
\end{abstract}


\end{frontmatter}


\section{Introduction}
Quantum entanglement is a vital concept for understanding many quantum information and computational tasks. Since the famous EPR  paper \cite{1}, entanglement of quantum states has attracted many scientists' attention. To understand this quantum phenomenon, fundamental features of entanglement have been studied \cite{2}. In the field of quantum information processing, bipartite entanglement has been better understood by its studying creation, quantification and manipulation \cite{2,3,4,5,6}. Using local operations, bipartite entangled states can be converted from one to another but it has been claimed that this is not possible for multipartite entangled states \cite{7}. However, recent findings show that transforming one class of multipartite entangled states to another is achievable under stochastic local operations and classical communication (SLOCC). Entanglement transformation between $W$ and $GHZ$ states have been shown, and the conversion rate between these states have been studied using the concepts of degeneration and border rank of tensors from algebraic complexity theory \cite{a}. In another work, it has been proven that obtaining a $W$ state from a $GHZ$ state with unit rate is possible, and transforming $GHZ$ states into $W$ states has also been shown \cite{b}. These methods are important since they open alternative ways of building a large-scale multipartite networks. $W$- and $GHZ$-state generation schemes can be compared regarding practical and experimental feasibility, and if the methods for the creation of the desired type are less feasible, one can prefer using a scheme to get the other type of state which is to be transformed into the desired type via SLOCC as discussed in \cite{a,b}. Because multipartite entangled states are required to implement some quantum information tasks, it is necessary to construct large-scale quantum-state networks. For example, multipartite entangled states are used for quantum teleportation \cite{8} and quantum key distribution \cite{9}. Also, some specific tasks require a particular type of multipartite entangled states, such as $GHZ$ states \cite{10}, for reaching a consensus in distributed networks and $W$ states are required for realizing an optimal universal quantum cloning machine \cite{11}. $GHZ$ and cluster states have been created \cite{12,13}, but because of its sophisticated structure $W$ states are relatively hard to create in large scales. Some theoretical and experimental reports have proposed expanding a $W$ state with ancillary photons or fusing two $W$ states \cite{14,15,16,17,18,19,20}. Ozdemir et al. \cite{21} succeeded in fusing two $W$ states to generate large scale $W$ states. Large scale $W$ states are important to construct high capacity information processors. They proposed an optical setup that fused two $W$ states of arbitrary sizes larger than or equal to 3. By integrating a Fredkin gate into this setup, a larger-scale resultant $W$ state was obtained with a higher probability \cite{22}. There are also theoretical proposals to fuse three and four $W$ states \cite{23,24} on optical setups including the basic fusion gate \cite{21}. As the size of the input $W$ states increased, the success probability of fusion decreased, increasing the resource cost for these setups. Lately, a new optical scheme has been proposed showing that it is possible to expand W states deteministically \cite{25}. This scheme doubles the size of an arbitrary W state deterministically by accessing locally all N qubits of the input W state. There are also protocols proposed for concentrating arbitrary less-entangled W state into a maximally entangled W state \cite{con1, con2}. Instead of expansion or concentration, we here perform creation operation to obtain a W state of arbitrary size, which is also a deterministic process. The current scheme is all-optical setup, and requires only single- and two-qubit gates. The successful application probabilities of these gates, which are not considered at this point, will be discussed in the last section. 


\section{Creation circuit for 3-, 4- and 5-qubit $W$ states} 
In this work, we propose quantum circuits that can be implemented to generate a $W$ state of any size using a certain number of non-entangled photons. First, we showed that three photons can be used as inputs to obtain a $W$ state of three qubits. Then, we increased our number of input photons by one to create a $W$ state of four qubits. After analyzing these two algorithms, we noticed that we could continue to enhance our $W$-state generating circuit to increase the size of the resultant W state. This led to the conclusion that we can use $n$ number of photons to generate an $n$-qubit $W$ state using only two-qubit gates. These gates are $F$ gates, previously shown to work in an optical setup \cite{25}, and $cNOT$ gates . The size of our resultant $W$ state depends on how many photons are used as input qubits. The number of photons is equivalent to the number of qubits of the resultant $W$ state. A $cNOT$ gate is a two-qubit gate that changes the polarization of the target qubit when the control qubit is vertically ($V$) polarized. The $F$ gate is also a two-qubit gate composed of 4 half-wave plates ($HWP$s) and a $cNOT$ gate. The $F$ gate acts on the target qubit when the control qubit is $V$ polarized. The action of the $F$ gate depends on how we arrange the $HWP$s on circuit. 
\begin{figure}[t!]
\includegraphics[width=1.0\textwidth]{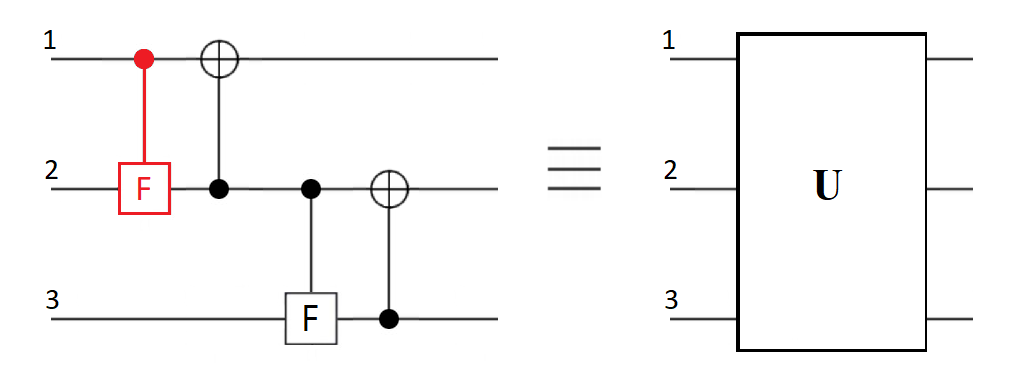}
\caption{Three photons are sent into the circuit consisting of two $F$ gates and two $cNOT$ gates. All of the two-qubit operations are shown in box $U$ for further use. The input state, $\ket{VHH}$ is transformed into a $W$ state of three qubits.} \label{fig:1}
\end{figure}
\begin{figure}[b!]
\includegraphics[width=1.0\textwidth]{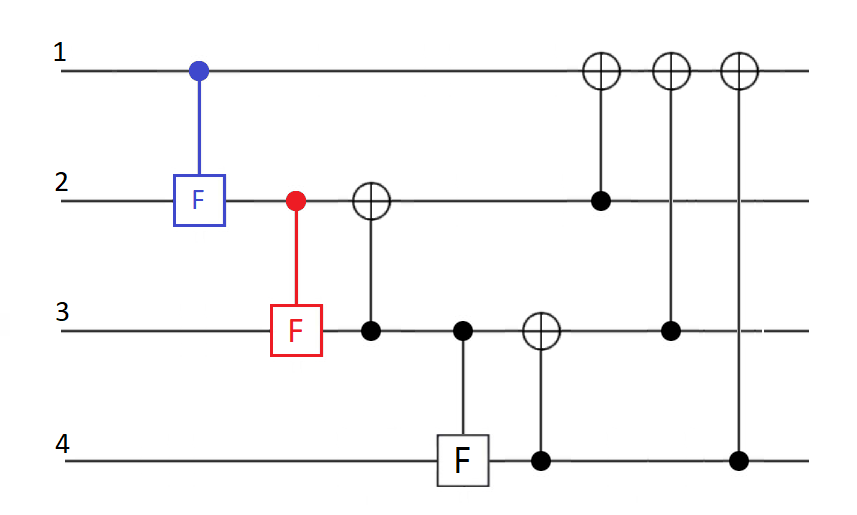}
\caption{The optical scheme which is the enhanced version of the three-qubit $W$-state creation circuit. This circuit is composed of eight two-qubit gates, three of which are $F$ gates. The other ones are $cNOT$ gates. Four photons are sent into the circuit to create a $W$ state of four qubits.} \label{fig:2}
\end{figure}
Now, let us look at the case where three photons are put into our circuit shown in Fig.\ref{fig:1}. There are two $F$ gates and two $cNOT$ gates. The input photon in spatial mode 1 is $V$ polarized whereas the other photons are horizontally ($H$) polarized. In each ket, the left qubit is in spatial mode 1 and the number of spatial modes increases from left to right. The $F$ gate in red transforms differently from the transformation of the $F$ gate in black as shown below. The black $F$ gate is a controlled Hadamard gate because it is composed of $HWP$s working at an angle of $\pi/8$. The $F$ gate is a generalized term for all angles. $cNOT$ gates are denoted by $cNOT$ and $F$ gates are denoted by $F$. Because the two-qubit gates act on different subgroups of the input photons, we put two subindices showing the spatial modes of the control and target qubit, respectively. All transformations of the gates including the $cNOT$s are as follows:
\begin{equation}
\begin{split}
 F_{23} cNOT_{21}  F_{12} & \ket{VHH} = F_{23} cNOT_{21} [ \ket{V} \otimes (\frac{1}{\sqrt{3}}\ket{H} + \sqrt{\frac{2}{3}}\ket{V}) \otimes \ket{H} ] \\
&= F_{23} cNOT_{21} [ \frac{1}{\sqrt{3}} \ket{VHH} + \sqrt{\frac{2}{3}} \ket{VVH} ] \\
&= F_{23} [ \frac{1}{\sqrt{3}} \ket{VHH} + \sqrt{\frac{2}{3}} \ket{HVH} ] \\
&= \frac{1}{\sqrt{3}} \big(\ket{VHH} + \ket{HVH} + \ket{HVV} \big). \\
\end{split}
\end{equation}
Finally, the last $cNOT$ gate acts on our state resulting in a $W$ state of three qubits,
\begin{equation}
\begin{split}
cNOT_{32} [ \frac{1}{\sqrt{3}} \big(\ket{VHH} + \ket{HVH} + \ket{HVV} \big) ] \\
= \frac{1}{\sqrt{3}} \big(\ket{VHH} + \ket{HVH} + \ket{HHV} \big).
\end{split}
\end{equation}
When we used four two-qubit gates we achieved creation of a $W$ state containing three qubits.

\begin{figure}[t!]
\includegraphics[width=1.0\textwidth]{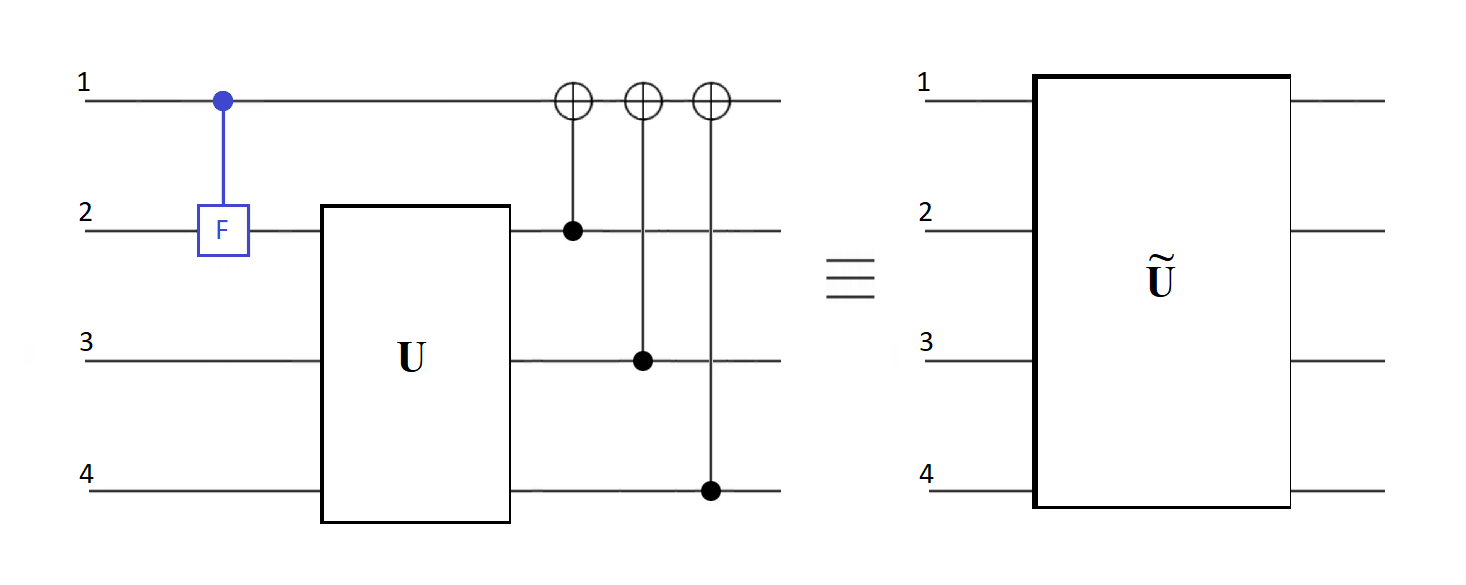}
\caption{The circuit used for creating a four-qubit $W$ state. By integrating three $cNOT$s and one $F$ gate, we enhance the circuit $U$ that gives us a three-qubit $W$ state. By enhancing the network we are able to create a larger-scale $W$ state by one qubit. The larger-scale $W$ state is a four-qubit state. The whole network is shown as $\tilde{U}$.} \label{fig:4}
\end{figure}

\begin{figure}[t!]
\centering
\includegraphics[width=0.8\textwidth]{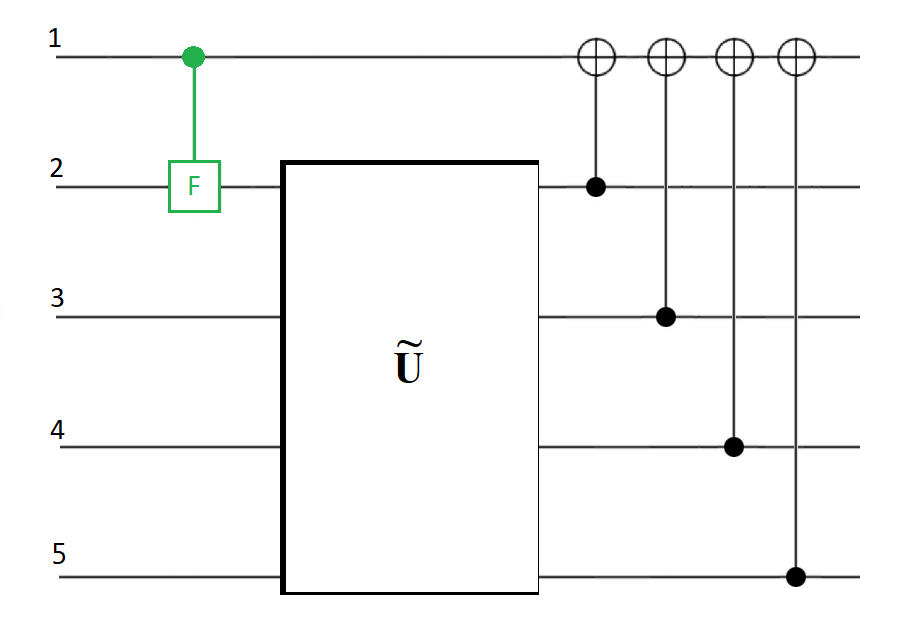}
\caption{The creation scheme for a five-qubit $W$ state. This circuit is composed of four $cNOT$ gates, a $F$ gate and $\tilde{U}$ shown in Fig.\ref{fig:4}. The number of gates added to the previous scheme, $\tilde{U}$, is equal to the number of qubits belonging to the resultant state.}
\label{fig:5}
\end{figure}

Now, let us look at the case where we use four photons to create a $W$ state of four qubits. The circuit consists of three $F$ gates and five $cNOT$s as shown in Fig. \ref{fig:2}. This circuit is shown with the $U$ box [Fig.\ref{fig:4}]. 
The action of the $U$ box is as follows: $U \ket{VHH} = \ket{W_{3}}$ where $\ket{W_{3}}$ is a three-qubit $W$ state. Let us show our operations step by step:
\begin{equation}
F_{12} \ket{VHHH} = \frac{1}{\sqrt{4}} \ket{VHHH} + \sqrt{\frac{3}{4}} \ket{VVHH}.
\end{equation}
$U$ box acts only on the second component of the resultant state because the photon in the spatial mode 2 is $\ket{V}$ and so
\begin{equation}
\begin{aligned}
U F_{12} \ket{VHHH} &= U \big[\ket{V} \otimes \big(\frac{1}{\sqrt{4}} \ket{HHH} + \sqrt{\frac{3}{4}} \ket{VHH}\big) \big] \\
&= \ket{V} \otimes \big(\frac{1}{\sqrt{4}} \ket{HHH} + \sqrt{\frac{3}{4}} \ket{W_{3}}\big)\\
&= \frac{1}{\sqrt{4}} (\ket{VHHH} + \ket{VVHH} + \ket{VHVH} + \ket{VHHV}).
\end{aligned}
\end{equation}
The last three $cNOT$ gates lead to a $W$ state of four qubits, which is
\begin{equation}
\begin{aligned}
cNOT_{21} cNOT_{31}& cNOT_{41}[\frac{1}{\sqrt{4}} (\ket{VHHH} + \ket{VVHH} + \ket{VHVH} + \ket{VHHV})] \\
&= \frac{1}{\sqrt{4}} (\ket{VHHH} + \ket{HVHH} + \ket{HHVH} + \ket{HHHV}) \\
&= \ket{W_{4}}.
\end{aligned}
\end{equation}
By adding an extra photon to the system again as shown in Fig. \ref{fig:5}, we can obtain a five-qubit $W$ state.
The previous network, which is used to create a $W$ state of four qubits, is shown as $\tilde{U}$. The input photons are transformed as follows:
\begin{equation}
\tilde{U}  F_{12} \ket{VHHHH} = \frac{1}{\sqrt{5}} \ket{VHHHH} + \sqrt{\frac{4}{5}} \ket{V} \otimes \ket{W_{4}},
\end{equation}
\begin{equation}
\begin{aligned}
&cNOT_{21} cNOT_{31} cNOT_{41} cNOT_{51} \tilde{U} F_{12} \ket{VHHHH} \\
&= cNOT_{21} cNOT_{31} cNOT_{41} cNOT_{51} \big[\frac{1}{\sqrt{5}} \ket{VHHHH} + \sqrt{\frac{4}{5}} \ket{V} \otimes \ket{W_{4}}\big] \\
&= \frac{1}{\sqrt{5}} (\ket{VHHHH} + \ket{HVHHH} + \ket{HHVHH} + \ket{HHHVH} + \ket{HHHHV}) \\
&= \ket{W_{5}}.
\end{aligned}
\end{equation}


\section{Generalized formula for the number of gates used}
We have constructed three circuits using 4 two-qubit gates, 8 two-qubit gates and 13 two-qubit gates to obtain three, four and five qubit $W$ states, respectively. When we want to increase the number of qubits of our $W$ state by one qubit, we integrate extra $n$ two-qubit gates where $n$ is the size of the resultant $W$ state. By performing straightforward calculations, the generalized formula for the number of gates used is
\begin{equation}
\frac{n(n+1) - 4}{2}, 
\end{equation}
where $n$ is the size of the resultant $W$ state. The number of $F$ gates used is $n-1$ whereas the number of $cNOT$ gates used is $\frac{(n-2)(n+1)}{2}$. To create a $W$ state of $n$ qubits, $n-1$ $F$ gates are used, with transformations as follows:
\begin{equation} \label{eq:HWP}
\begin{array}{cc}
F_{(n-1) n}\ket{10} = \ket{1} \otimes \frac{1}{\sqrt{2}}(\ket{0} + \ket{1}), \\
F_{(n-2) (n-1)}\ket{10} = \ket{1} \otimes (\frac{1}{\sqrt{3}}\ket{0} + \sqrt{\frac{2}{3}}\ket{1}),\\
.\\
.\\
.\\
F_{23} \ket{10} = \ket{1} \otimes (\frac{1}{\sqrt{n-1}}\ket{0} + \sqrt{\frac{n-2}{n-1}}\ket{1}),\\
F_{12} \ket{10} = \ket{1} \otimes (\frac{1}{\sqrt{n}}\ket{0} + \sqrt{\frac{n-1}{n}}\ket{1}),
\end{array}
\end{equation}
where the subindices show the spatial modes of the control and target qubits respectively.
One can also look at Tab.\ref{tab:1} to see how many gates we used for the creation of certain $W$ states.

\begin{table}[t!]
\centering
\caption{\label{tab:1} The number of gates used for the creation of certain $W$ states.}
\begin{tabular}{cccc}
\hline 
  Resultant $W$ state &  Two-qubit gates used &  $F$ gates &  $cNOT$ gates   \\
 \hline
    $\ket{W_{3}}$ & 4 & 2 & 2 \\
    $\ket{W_{4}}$ & 8 & 3 & 5 \\ 
    $\ket{W_{5}}$ & 13 & 4 & 9 \\ 
    $\ket{W_{6}}$ & 19 & 5 & 14 \\ 
    $\ket{W_{7}}$ & 26 & 6 & 20 \\ 
\hline
\end{tabular}
\end{table}

The total number of two-qubit gates needed to obtain a $W$ state of $n$ qubits is proportional to $n^2$, which means that it increases quadratically.


\section{Decomposition of an $F$ gate and implementation of the circuit}
\begin{figure}[t!]
\centering
\includegraphics[width=1.0\textwidth]{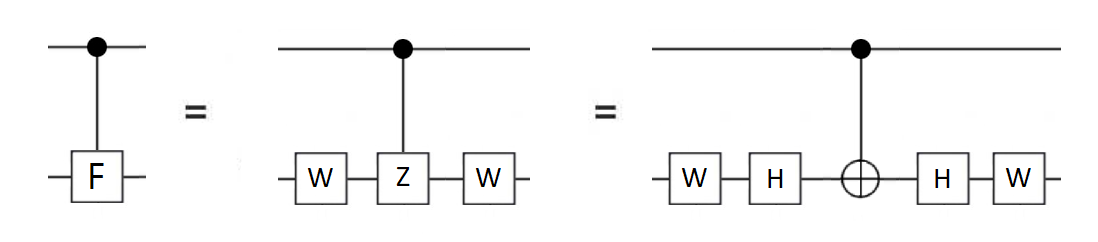}
\caption{Decomposition of the $F$ gate into 4 $HWP$s and a $cNOT$ gate. $W$ gates are $HWP$s working at a certain angle that determines the transformation of a specific $F$ gate. Hadamard gates can be implemented using $HWP$s working at the angle $\pi/8$. But $HWP$s are not equivalent to a Hadamard gate because of its nonunitary nature. Therefore we use $HWP$s to perform Hadamard operation.} \label{fig:6}
\end{figure}

Previously, we defined how $F$ gates act. Each of these gates can be decomposed into three optical elements that are two $HWP$s working at some rotation angle, $\theta$, and a $cZ$ gate. Also, a $cZ$ gate can be decomposed into a $cNOT$ gate and two $HWP$s acting as Hadamard gates. So, the total number of two-qubit gates needed is equivalent to the number of $cNOT$s that are to be used. In Fig. \ref{fig:6}, the optical circuit corresponding to the $F$ gate is shown, and their matrix representations are as follows:
\begin{equation}
W = \left( \begin{matrix} \cos{\theta}&\sin{\theta} \\ \sin{\theta}&-\cos{\theta}\end{matrix} \right),\\ cZ = \left( \begin{matrix} 1&0&0&0 \\ 0&1&0&0 \\ 0&0&1&0 \\ 0&0&0&-1\end{matrix} \right).
\end{equation}
Each transformation of the $F$ gates depends on the angle at which the half-wave plates manipulate the polarization of the photons. The general formula of a $F$ gate is:
\begin{equation}
F = \left( \begin{matrix} 1&0&0&0 \\ 0&1&0&0 \\ 0&0&\cos{2\theta}&\sin{2\theta} \\ 0&0&\sin{2\theta}&-\cos{2\theta}\end{matrix} \right)
\end{equation}
where the sine and cosine elements determine the factors of the components of the output state.

\begin{figure}[b!]
\centering
\includegraphics[width=0.7\textwidth]{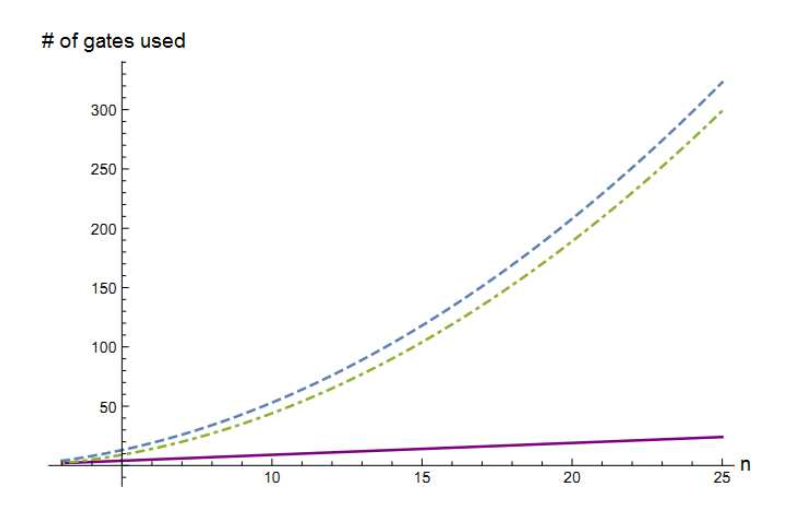}
\caption{The blue dashed line shows the relationship between the total number of two-qubit gates required and the number of qubits belonging to the resultant $W$ state. The green dashed-dotted line shows the relationship between the number of $cNOT$s and $n$, the number of qubits of $W$ state. The purple solid line shows the relationship between $n$ and the number of $F$ gates used.} \label{fig:8}
\end{figure}

A graph showing the relationship between the number of qubits $n$ and the total number of $cNOT$s can be seen in Fig. \ref{fig:8}.

\begin{figure}[b!]
\centering
\includegraphics[width=0.8\textwidth]{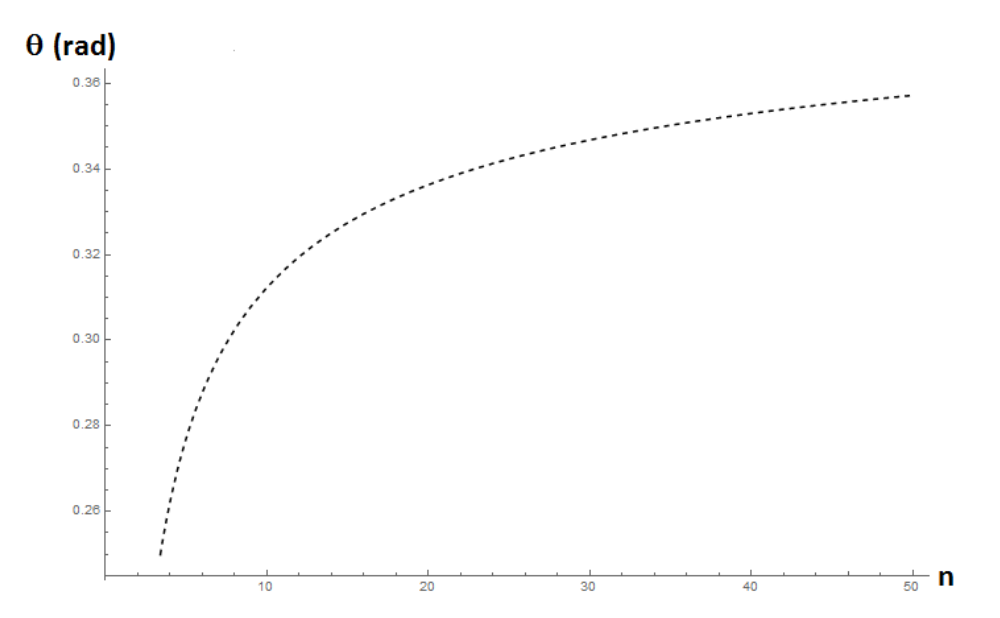}
\caption{A graph showing the relationship between the size of the resultant $W$ state and the angle at which the first $HWP$ acts. All of the other $HWP$s act as shown in Eq.\ref{eq:HWP}. The relationship between the angles of the other $HWP$s and $n$ can also be seen in the graph.  This graph shows how we should arrange our $HWP$s such that the circuit generates a $W$ state of $n$ qubits.} \label{fig:7}
\end{figure}

The relationship between the size of the resultant $W$ state and the angle of the first $HWP$ is shown in Fig. \ref{fig:7}. The mathematical relation is
\begin{equation}
\theta = \frac{1}{4} \arccos [\frac{1}{\sqrt{n}}].
\end{equation}

\section{The realization of the proposed schemes}

Although the proposed schemes allow us to create any desired W state, we lack ideal conditions regarding experimental realization for the sophisticated structure of these methods. One of the drawbacks of our setup is that as the size of $W$ state gets larger, the number of required gates increases quadratically. Therefore, this leads to a large number when we would like to create large-scale $W$ states. For large $n$, we have shown that the number of $cZ$ gates is $\propto{n}$. This value is very small compared with the number of $cNOT$ gates that is $\propto{n^2}$. Because the action of the $cZ$ gate corresponds to the operation done by two $HWP$s and a $cNOT$ gate, we need to take into account the implementation of the $cNOT$ gate. Here, we emphasize the importance of the experimental implementation of the $cNOT$ gate by showing that the creation of large-scale $W$ states can be done using only two-qubit gates. There are theoretical and experimental proposals for the implementation of the $cNOT$ gate. Also, the $cNOT$ gate has been experimentally shown to work with a probability of $1/9$ in linear optics \cite{27}. This means that the probability of the successful creation of an $n$-qubit $W$ state is $({1/9})^\frac{n(n+1) - 4}{2}$. Successful creation probability is very low, even for a three-qubit $W$ state with a success probability of the order of $10^{-4}$. However, for the last decade, there have been proposals to implement optical gates using Kerr nonlinearities \cite{extra2,extra3,26}. In one of these, Nemoto et al. \cite{26} pointed out that weak cross-Kerr nonlinearities are more useful to construct a $cNOT$ gate with fewer physical resources than other linear optical schemes. Therefore, this work increases the importance of cross-Kerr nonlinearities and its application for $cNOT$ gates. There are also other methods to realize a $cNOT$ gate using superconductors \cite{28} and ion-trap systems \cite{29,30}.

Another experimental imperfection is that, for $n$ is very large, the process may lead to failure case since a few degrees of deviation in angle changes the outcome, which means desired $W$ state is not created. For example, to create a $W$ state of $100$ qubits, we need to set the angle at $22.05^{\circ}$ for the first $HWP$. If we accidentally set the angle $21.5^{\circ}$, we get a failure case because this angle is approximately the angle required for the first $HWP$ to get a $W$ state of $200$ qubits. This small change in the angle changes the output greatly, leading to failure for creation process. Theoretically, we here assume that the angle setting is adjusted as required for a $W$-state of $n$ qubits. The precision of angles will be a challenging issue for experimentalists. Also, the effect of photon loss (the amplitude damping channel) may lead to experimental imperfections as discussed in \cite{extra}. 

We should also take into account the practical source of single photons to consider the feasibility of the proposed schemes. Parametric 
down-conversion (PDC) is a common and well-known method to generate single-photon states, and has been discussed for the scheme expanding polarization entangled $W$ states \cite{15}. Suppose that single photons are generated from PDC with rate $\gamma$, Three single photons required for the preparation of a three-qubit $W$ state are generated with rate ${\gamma}^3$. In this case, the undesirable events usually occur due to the generation of an extra photon pair, with rate $\delta$ which is $\sim 10^{-4}$. Such events occur with rate $O({\gamma}^3 \delta)$, which is very small compared with the rate $O({\gamma}^3)$. It can be generalized to the case of n-photon $W$-state creation, which gives the rates of the desired events and errors, $O({\gamma}^n)$ and $O({\gamma}^n \delta)$ respectively. As in the simplest example of three-photon $W$-state creation, the rate of errors is small compared with the rate of the desired events.

\section{Conclusion}

We have presented three optical setups that create $W$ states of three, four and five qubits deterministically. We then showed that a $W$ state of any size can be created deterministically by enhancing the network. We also derived the formula for the number of gates needed for this processing. An important capability of our proposal is that it can be used to obtain any $W$ state using a number of gates depending on the size of the desired resultant $W$ state. Also, we only need two-qubit gates, $cZ$ and $cNOT$, making our setup more applicable because we do not use any three- or more-qubit gates. The implementation of the $cNOT$ gate has been well characterized in the literature. It has been shown to be realized in both theoretical and experimental works. Although there are some challenges to overcome, for example the low success probability in the experimental realization of the $cNOT$ gate, our circuit can be realized with current technology.

\section{Acknowledgements}
I would like to thank Zafer Gedik for his useful discussions.

\end{document}